\documentclass[11pt,a4paper]{iopart}
\usepackage[utf8]{inputenc}
\expandafter\let\csname equation*\endcsname\relax

\expandafter\let\csname endequation*\endcsname\relax 
\usepackage{amsmath}
\usepackage{amsfonts}
\usepackage{amssymb}
\usepackage{makeidx,bm}
\usepackage{graphicx}
\usepackage{iopams}
\usepackage{todonotes}
\usepackage{hyperref}

\usepackage{soul}

\def\bea{\begin{eqnarray}}
\def\eea{\end{eqnarray}}
\def\n{\nonumber}

\def\la{\langle}
\def\ra{\rangle}

\begin{document}

\title{Effect of stochastic resetting on Brownian motion with stochastic diffusion coefficient}

\author{Ion Santra$^{1}$, Urna Basu$^{1,2}$, Sanjib Sabhapandit$^{1}$}

\address{$^{1}$Raman Research Institute, Bengaluru 560080, India}

\address{$^{2}$S. N. Bose National Centre for Basic Sciences, Kolkata 700106, India}

\begin{abstract}
We study the dynamics of a Brownian motion with a diffusion coefficient which evolves stochastically. We first study this process in arbitrary dimensions and find the scaling form and the corresponding scaling function of the position distribution. We find that the tails of the distribution have exponential tails with a ballistic scaling. We then introduce the resetting dynamics where, at a constant rate, both the position and the diffusion coefficient are reset to zero. This eventually leads to a nonequilibrium stationary state, which we study in arbitrary dimensions. In stark contrast to ordinary Brownian motion under resetting, the stationary position distribution in one dimension has a logarithmic divergence at the origin. For higher dimensions, however, the divergence disappears and the distribution attains a dimension-dependent constant value at the origin, which we compute exactly. The distribution has a generic stretched exponential tail in all dimensions.
 We also study the approach to the stationary state and find that, as time increases, an inner core region around the origin attains the stationary state, while the outside region still has a transient distribution---this inner stationary region grows $\sim t^2$, i.e., with a constant acceleration, much faster than ordinary Brownian motion.
\end{abstract}

\title[Stochastic resetting on Brownian motion with stochastic diffusion coefficient]
\maketitle

\section{Introduction}
Stochastic resetting refers to random intermittent interruptions and restarts of a dynamic process~\cite{resettingRev}. Since the seminal work of Evans and Majumdar~\cite{evans2011diffusion} a decade ago, stochastic resetting has made a profound impact in the field of nonequilibrium statistical physics because of its very rich features---attainment of a nonequilibrium stationary state (NESS) at long times~\cite{evans2011diffusion,evans2014diffusion}, optimization of search times~\cite{evans2011optiman,kusmierz2014first}, dynamical transition in relaxation to the stationary states~\cite{majumdar2015dynamical}, unusual thermodynamic properties~\cite{fuchs2016stochastic,pal2017integral}, and universal extreme value statistics~\cite{majumdar2021record,singh2021extremal}. Applications of stochastic resetting have also been studied in the context of biology~\cite{roldan2016stochastic}, ecology~\cite{bhat2016stochastic}, epidemiology\cite{ramoso2020stochastic}, and economics~\cite{santra2022effect}. The effect of stochastic resetting has been explored in a variety of stochastic systems including underdamped diffusions~\cite{gupta2019stochastic}, L\'evy walks and L\'evy flights~\cite{kusmierz2014first,kusmierz2015optimal}, random acceleration process~\cite{singh2020random}, fractional Brownian motion~\cite{majumdar2018spectral}, active particles~\cite{evans2018run,santra2020run,kumar2020active} and several others~\cite{goswami2021stochastic,mercado2021search,zanette2020fat}.

 Brownian motion with a stochastically evolving diffusion coefficient has gained considerable interest in the past few years~\cite{chubynsky2014diffusing,
 chechkin2017brownian,jain2016diffusion,tyagi2017non,sposini2020universal,wang2022random}. Such models were first proposed to explain the `Fickian yet non-Gaussian’ diffusion of tracers in hard-sphere colloidal suspensions, nanoparticles in nanopost arrays, fluid interfaces and moving along membranes,
and nematode motions~\cite{ddEg1,ddEg2,ddEg3,ddEg4,ddEg5}. Random diffusion constants are also ubiquitous in financial mathematics, commonly known as stochastic volatility models~\cite{durlauf2016macroeconometrics}. The simplest  scenario of a Brownian motion $x(t)$ with a stochastically evolving diffusion coefficient $D(t)$ is~\cite{sposini2020universal},
\begin{align}
\frac{dx}{dt}&=\sqrt{2D(t)}\,\eta(t),\label{intro1}\\
\intertext{where}~ D(t) &= \omega^2(t) \quad\text{with} \quad
{d\omega\over dt} =\sqrt{2\Lambda^2}\, \zeta(t).
\label{model:intro}
\end{align}
Here $\Lambda$ is a constant and $\eta(t)$ and $\zeta(t)$ are independent white noises with autocorrelations $\la \eta(t)\eta(t') \ra=\la \zeta(t)\zeta(t') \ra=\delta(t-t')$ and $\la \eta(t)\zeta(t') \ra=0$.
 Such stochastic diffusion coefficient $D(t)$ would appear in a medium undergoing heating, where Gaussian fluctuations are expected about a mean temperature that increases with time. A similar process also arises naturally in the context of direction reversing active Brownian particles~\cite{santra2021active}, often used to model the motion of a class of bacteria like M. xanthus and P. putida~\cite{thutupalli2015directional}. Unlike the ordinary Brownian motion, the stochastically evolving diffusion coefficient leads to the position variance $\la x^2(t) \ra\propto t^2$ as well as strongly non-Gaussian position fluctuations. A natural extension is to study the effect of stochastic resetting on systems described by~\eref{model:intro}.

In this paper, we study the effect of resetting on a $d$-dimensional generalization of the Brownian motion defined by~\eref{intro1},
\begin{align}
\frac{dx_n}{dt}=\sqrt{2D(t)}\,\eta_n(t), \quad n=1,2,\dotsc,d,
\label{ddim:Langevin}
\end{align}
where $\la \eta_i(t)\eta_j(t') \ra=\delta_{ij}\,\delta(t-t')$. Note that all the components of the position $\{x_1(t),x_2(t),\dotsb,x_d(t)\}$ share a common diffusion coefficient $D(t)$, evolving according to \eref{model:intro}. First, we show that, in the absence of resetting, the probability distribution of the radial coordinate $r=\sqrt{ x_1^2+x_2^2+\dotsb+x_d^2}$ admits a scaling form,
\begin{align}
P(r,t)=\frac{1}{4\Lambda t}f_d\Bigg(\frac{r}{4\Lambda t}\Bigg).
\end{align}
We also show that the scaling function has a universal exponential tail  $f_d(z)\sim e^{-\pi z}$, where the dependence on dimensions appears only through the subleading prefactor [see \eref{fr-asym}].

The aim of this paper is to explore the effect of resetting on the above process. We study the resetting protocol where both the position and the diffusion coefficient of the particle are stochastically reset to zero at a constant rate $\alpha$. We find that the position eventually reaches a nonequilibrium stationary state, which we characterize analytically. In particular,  the radial distribution has the scaling form,
\begin{align}
P_\alpha^s(r)=\frac{\alpha}{4\Lambda}H_d\Big(\frac{\alpha r}{4\Lambda} \Big),
\end{align}
where the scaling function has a universal stretched exponential tail $H_d(w)\sim e^{-2\sqrt{\pi w}}$. The dependence on dimensions appears only through the subleading prefactor [see \eref{tails_bessel} and \eref{tails_ddim}]. On the other hand, the behavior of $H_d(w)$ as $w\to 0$ for $d=1$ is very different from that for $d>1$: it has a logarithmic  divergence for $d=1$, whereas it approaches a $d$-dependent finite value for $d>1$. 
We also find that the relaxation to the stationary state follows a similar mechanism as described in~\cite{majumdar2015dynamical}, where there is an inner stationary region and an outer transient region. In our case, the stationary region expands with a constant acceleration, as opposed to a constant velocity expansion in the case of ordinary Brownian motion~\cite{majumdar2015dynamical}.

 The paper is organized as follows. We first consider the position distribution for the one-dimensional case in the absence of resetting in Sec.~\ref{sec:noreset1d}. The position distribution for the general $d$-dimensional process without resetting is obtained in Sec.~\ref{sec:noresetdd}. 
  The stationary position distribution in the presence of resetting is discussed in Sec.~\ref{sec:R1d} and Sec.~\ref{sec:Rdd} for $d=1$ and $d\geq 2$, respectively. The relaxation to the stationary state is studied in 
 Sec.~\ref{s:approach}. Finally, we summarize and conclude in Sec.~\ref{concl}.

\section{Position distribution in one dimension}\label{sec:noreset1d}
The position distribution for the one-dimensional Brownian motion with a stochastic diffusion coefficient defined by equations \eref{intro1} and \eref{model:intro} was obtained in~\cite{sposini2020universal}. In this section, we re-derive the same using a different method, which we use later for generalization to arbitrary dimensions.

We start with the Langevin equations \eref{intro1} and \eref{model:intro} with the initial condition $x(0)=0=\omega(0)$.
The corresponding Fokker-Planck equation for the joint distribution $\mathsf{ P}(x,\omega,t)$ can be written down as,
\begin{equation}
\frac{\partial \mathsf{ P} }{\partial t} =\omega^2\frac{\partial^2 \mathsf{ P}}{\partial x^2}+ \Lambda^2\frac{\partial^2 \mathsf{ P}}{\partial \omega^2}.
\label{fpeq}
\end{equation}
The initial and boundary conditions for \eref{fpeq} are respectively, $\mathsf{P}(x,\omega,0)=\delta(x)\delta(\omega)$, and $\mathsf{ P}(x,\omega,t)\to 0$ for both $x\to\pm \infty$ and  $\omega\to\pm \infty$. To solve \eref{fpeq}, it is convenient to consider the Fourier and the Laplace transform with respect to $x$ and $t$ respectively, defined by,
\begin{align}
\tilde{\mathsf{P}}(k,\omega,s)=\int_{0}^{\infty}dt \, e^{-st} \int_{-\infty}^\infty dx \, e^{ikx} \mathsf{P}(x,\omega,t).
\end{align}
Upon performing this transformation on \eref{fpeq}, we obtain an ordinary second order differential equation for $\tilde{\mathsf{P}}(k,\omega,s)$,
\begin{equation}
\left[ \Lambda^2\frac{d}{d \omega^2}  -\bigl(s+k^2 \omega^2\bigr)\right] \tilde{\mathsf{P}}(k,\omega,s) = -\delta(\omega),
\label{fpftlt}
\end{equation}
with the boundary conditions $\tilde{\mathsf{P}}(k,\omega,s) \to 0$ for $\omega\to\pm\infty$. For $\omega\ne 0,$ the general solution of \eref{fpftlt} is given by,
\begin{align}
\tilde{\mathsf{P}}(k,\omega,s)&=a~ \mathbb D_{-q}\left(\omega \sqrt{\frac{2k}\Lambda}\right)+b~\mathbb D_{-q}\left(-\omega \sqrt{\frac{2k}\Lambda}\right)\quad \text{with}~~q=\frac 12 \Big(1 +\frac{s}{k\Lambda}\Big),
\end{align}
where $a,b$ are arbitrary $\omega$-independent constants and $\mathbb D_\nu(z)$ denotes the parabolic cylinder function \cite{dlmf}.

 Using the boundary conditions for $\omega \to\pm\infty$, and the fact that $\tilde{\mathsf{P}}(k,\omega,s)$ is continuous at $\omega=0$  we have,
\begin{align}
\tilde{\mathsf{P}}(k,\omega,s)= a~\mathbb D_{-q}\left(|\omega| \sqrt{\frac{2k}\Lambda}\right).
\label{homosol}
\end{align}
Integrating \eref{fpftlt} over $\omega$ from $\omega=-\epsilon$ to $\omega=\epsilon$ and taking the limit $\epsilon\to 0$, yield the discontinuity in the first derivative of $\tilde{\mathsf{P}}(k,\omega,s)$ at $\omega=0$,  
\begin{align}
\frac{d \tilde{\mathsf{P}}}{d \omega}\bigg|_{\omega=0^+}-\frac{d \tilde{\mathsf{ P}}}{ d\omega}\bigg|_{\omega=0^-}=-\frac{1}{\Lambda^2}.
\end{align}
We use this discontinuity to determine $a$, which finally yields,
\begin{align}
\tilde{\mathsf{P}}(k,\omega,s)&= \frac{2^{\frac q 2}}{\sqrt{8 \pi k \Lambda^3}} \Gamma\left(\frac q 2 \right) \mathbb D_{-q}\left(|\omega| \sqrt{\frac{2k}\Lambda}\right).
\end{align}
Since we are interested in the position distribution, we integrate over $\omega$ to get,
\begin{align}
\tilde{P}(k,s)=\frac{2}{s + k\Lambda} ~  _2F_1\left(1,\frac{q+1}2,\frac {q+2}2,-1\right),
\end{align}
where $\tilde{P}(k,s)$ denotes the Fourier-Laplace transform of the position distribution $P(x,t)$ and $_2F_1(a,b,c,z)$ denotes the Hypergeometric function \cite{dlmf}. 
The Laplace transformation can be inverted exactly (see Sec. V in the Supplemental material of~\cite{santra2021active}), which leads to the characteristic function of the position distribution,
\begin{align}
\hat{P}(k,t)=\bigl\langle e^{i k x}\bigr\rangle  = {1\over \sqrt{\cosh \bigl(2 \Lambda k t  \bigr)}}.
\label{characteristic-function}
\end{align}
The distribution in real space can be obtained by taking the inverse Fourier transform of the characteristic function,
\begin{align}
P(x,t)=\frac{1}{2\pi}\int_{-\infty}^{\infty}dk\,{e^{-ikx}\over \sqrt{\cosh \bigl(2 \Lambda k t  \bigr)}}.
\end{align}
From the above expression, it is evident that the position distribution $P(x,t)$ is a function of the scaled variable $x/(\Lambda t)$. This ballistic scaling is in sharp contrast with the diffusive scaling $x/\sqrt{Dt}$ observed for ordinary Brownian motion, where $D(t)=D$ is a constant.
 
 A posteriori, the scaling form  
 \begin{equation}
P(x,t)={1\over 4\Lambda t}f\left({x\over  4\Lambda t}\right),
\label{pxt_scaled}
 \end{equation}
turns out to be more convenient, where the scaling function $f(y)$ is given by
\begin{equation}
f(y)=\int_0^\infty {d\kappa\over \pi} {\cos (\kappa y) \over\sqrt{\cosh (\kappa/2)}}.
\label{fy-integral}
\end{equation}
The above integral can be performed exactly and yields,
\begin{equation}
f(y)={1\over \sqrt{2} \,\pi^{3/2}}\,
\Gamma\left({1\over 4} +iy\right)
\Gamma\left({1\over 4} -iy\right),
\label{scaled-propagator}
\end{equation}
where $\Gamma(z)$ is the gamma function.

From the property of the gamma function, $\Gamma(z^*)=\Gamma^*(z)$---where $^*$ indicates the complex conjugate---\eref{scaled-propagator} immediately implies that $f(y)$ is a symmetric real-valued function. Moreover, it is normalized to unity, i.e., $\int_{-\infty}^\infty f(y)\, dy=1$, as required.  For large $|y|$, asymptotically, $f(y)$ decays with exponential tails,
\begin{align}
 f(y)\sim \sqrt{\frac 2\pi}\,  |y|^{-1/2}\,e^{-\pi |y|}.
 \label{free:largey}
\end{align} 

It is noteworthy that, although $\la D(t)\ra$ grows linearly with $t$ [following from \eref{model:intro}], the stochastic fluctuations of $D(t)$  make the position distribution \eref{pxt_scaled} very different from the case where $D(t)$ has deterministic linear growth with time. In fact, any deterministic $D(t)$ in \eref{model:intro} leads to a Gaussian distribution with the variance $\sigma^2(t)=2\int_0^t D(t')dt'$. 

In higher dimensions, for any deterministic $D(t)$, each position component $x_n(t)$ evolves independently. Therefore the $d$-dimensional position distribution is a product of Gaussian distributions of each component. This remains true even when each component $x_n(t)$ evolves with a different deterministic function $D_n(t)$.

 For stochastic diffusion coefficients, however, the generalization of \eref{intro1} to higher dimensions is not unique. In a trivial generalization, where each component $x_n(t)$ evolves with an independent stochastic diffusion coefficient $D_n(t)$, the $d$-dimensional position distribution is merely the product of the position distributions \eref{pxt_scaled} of each component. The more interesting case is where the components $\{x_n(t)\}$ evolve with a common stochastic $D(t)$, which leads to a non-factorized position distribution, as we see below.
 
\section{Position distribution in $d$-dimensions}\label{sec:noresetdd}
We start with the Langevin equations \eref{ddim:Langevin} for position component $x_n(t)$, where the common diffusion coefficient $D(t)$ evolves by \eref{model:intro}. We consider the initial condition $x_n(0)=0$ and $\omega(0)=0$, so that,
\begin{align}
x_n(t)=\int_0^{t} \sqrt{2D(t')}\,\eta_n(t')\,dt'.
\label{int:langevin}
\end{align}
The common diffusion coefficient make the different position components non-trivially correlated. In particular, the correlations between the even powers of the different position components are non-zero. For example, 
\begin{align}
\la x_n^2(t_1) x_m^2(t_2)\ra-\la x_m^2(t_1)\ra\la x_n^2(t_2)\ra=\frac{16}{3}\Lambda^4 t_2^3\,(2t_1-t_2),\quad\text{for}~~n\neq m,~~t_1>t_2.
\end{align}

The presence of non-vanishing correlations indicate that the position distribution $P(\mathbf{x},t)$ of the position vector $\mathbf{x}=[x_1,x_2,\dotsc,x_d]$ cannot be expressed in a factorized form.
To compute the distribution $P(\mathbf{x},t)$, it is convenient to consider the characteristic function,
\begin{align}
\tilde{F}(k)=\int d\mathbf{x}\,e^{i\mathbf{k}.\mathbf{x}}\,P(\mathbf{x},t)&=\langle e^{i\mathbf{k}.\mathbf{x}}\rangle   
=  \Bigl\langle  \prod_{n=1}^d  \exp\Big(i k_n \int_0^t \sqrt{2D(t')}\,\eta_n(t')\, dt'\Big) \Bigr\rangle, 
 \label{char3}
\end{align}
where $\bf{k}=[k_1,k_2,\dotsc,k_d]$ and we have used \eref{int:langevin}. The average $\langle \cdot\rangle$ is over both $\{\eta_n(t)\}$ and $\{D_n(t)\}$. For a given realization of $D(t)$, the average over the independent white noises $\{\eta_n(t)\}$ can be performed using the Gaussian property, resulting in
\begin{equation}
\tilde{F}(k)= \Bigl\langle \prod_{n=1}^d \exp\left(-k_n^2 \int_0^t D(t')\, dt'\right)\Bigr\rangle=\Bigl\langle \exp\left(-k^2 \int_0^t D(t')\, dt'\right)\Bigr\rangle,
\label{char4}
\end{equation}
 where the average over the stochastic trajectories of $\{D(t)\}$ is yet to be performed and
\begin{equation}
 k^2=\sum_{n=1}^d k_n^2=\mathbf{k}\cdot\mathbf{k}.
 \label{k2}
\end{equation}
 The right hand side of \eref{char4} is nothing but the characteristic function of the one-dimensional case, obtained in \eref{characteristic-function}. Therefore,
\begin{equation}
\tilde{F}(k) ={1\over \sqrt{\cosh(2k\Lambda t)}}.
\label{tildeFk}
\end{equation}
Since the right-hand side of the above equation depends only on the magnitude of $\mathbf{k}$, the distribution of $\mathbf{x}$ is isotropic, i.e., it only depends on the radial distance $r=\sqrt{x_1^2+x_2^2+\dotsb+x_d^2}$. 
For any isotropic probability distribution, $P(\mathbf{x}) \,d\mathbf{x}=F_d(|\mathbf{x}|)\,d\mathbf{x}$ is related to its Fourier transform $\tilde{F}(k)$ (which is also isotropic and depends only on $k=|\mathbf{k}|$) by the Hankel transform~\cite{dlmf},
\begin{equation}
r^{{d\over 2}-1} \, F_d(r)={1\over (2\pi)^{d/2}}\,
\int_0^\infty k^{{d\over 2}-1}\, \tilde{F}(k)\, J_{{d\over 2}-1}(k r) \, k\,dk.
\label{Fr}
\end{equation}
Note that we can also express $\tilde{F}(k)$ in terms of $F_d(r)$ in a similar fashion using the orthogonality of Bessel functions, 
$
\int_0^\infty J_\nu (k r) J_\nu(k' r) \, r\, dr=\delta(k-k')/k,\nonumber
$
for $\{k, k'\}>0.$

 Since $F_d(r)$ is the distribution in Cartesian coordinates, the radial distribution $\rho(r)$ is obtained by considering the appropriate volume element for $d$-dimensional spherical coordinates and integrating out the angular coordinates.  This is related to $F_d(r)$ by
\begin{equation}
\rho(r) = {2\pi^{d/2} r^{d-1}\over \Gamma(d/2)}\, F_d(r),
\label{fr}
\end{equation}
 where $\int_0^\infty \rho(r)\, dr=1$.
Therefore, from \eref{tildeFk}, we get the radial distribution as,
\begin{align}
\rho(r,t)={2 \,r^{d/2} \over 2^{d/2}\Gamma(d/2) } \int_0^\infty 
{k^{{d\over 2}-1} J_{\frac{d}{2}-1}(k r) \over \sqrt{\cosh (2k\Lambda t)} }\, k\,dk.
\end{align}
It is evident that $\rho(r,t)$ has the scaling form,
\begin{align}
\rho(r,t)=\frac{1}{4\Lambda t}f_d\Bigg(\frac{r}{4\Lambda t}\Bigg),
\label{radialfree}
\end{align}
where the scaling function is given by,
\begin{equation}
f_d(z)={2 \,z^{d/2} \over 2^{d/2}\Gamma(d/2) } \int_0^\infty 
{\kappa^{{d\over 2}-1} J_{\frac{d}{2}-1}(\kappa z) \over \sqrt{\cosh (\kappa/2)} }\, \kappa\,d\kappa.
\label{f_d-r}
\end{equation}
Evidently, $\int_0^\infty f_d(z) \, dz =1$. 
Using the recursive properties of Bessel function $J_\nu (x)$, the above equation can be also written as,
\begin{equation}
f_d(z)={\,z^{d-1}\over 2^{\frac{d-4}{2}}\Gamma(d/2)}\,
\left(-{1\over z} {d\over d z}\right)^n\left[{M_\beta(z)\over z^\beta}\right]\quad\text{with}~n=\frac{d-2\beta}{2}-1, 
 \label{f_d-r2}
\end{equation}
where $\beta=-1/2$ and $0$ respectively for odd and even $d$, and 
\begin{equation}
M_\beta(z) =\int_0^\infty {\kappa^{\beta} J_\beta(\kappa z) \over  \sqrt{\cosh (\kappa/2)} }\, \kappa\,d\kappa.
\label{G-alpha}
\end{equation}

For odd dimensions, i.e., $\beta=-1/2$, using $J_{-1/2}(x)=\sqrt{2/\pi} \,\cos(x)/\sqrt{x}$, the above integral can be performed explicitly, which gives $\sqrt{z} M_{-1/2}(z)=\sqrt{2\pi}\, f(z)$, where $f(z)$ is the one-dimensional scaled distribution, given by \eref{scaled-propagator}. Therefore, $f_d(z)$ from \eref{f_d-r2} can be expressed explicitly in terms of derivatives of $f(z)$  in all odd dimensions,
\begin{align}
f_d(z)={2 \,z^{d-1}\sqrt{\pi}\over 2^{(d-1)/2}\Gamma(d/2)}\,
\left(-{1\over z} {d\over d z}\right)^{\frac{d-1}{2}}f(z).
\label{ddim:oddexact}
\end{align}
 For example,  
\begin{math}
f_3(z)=-2 z f'(z),
\end{math}
for three dimensions.

For even dimensions, i.e., $\beta=0$, the integral for $M_0(z)$ in \eref{G-alpha} is difficult to evaluate exactly. However, the asymptotic behavior for large $z$ can be systematically extracted. In general, for any $d$ (odd or even), the scaled radial distribution has a universal exponential tail [see \ref{App:A}], 
\begin{equation}
 f_d(z) \approx \frac{2\sqrt{\pi}}{\Gamma(d/2)}\left(\frac{\pi z}{2}\right)^{\frac d2-1}  \,e^{-\pi z}
\quad\text{for}~z\gg 1.
\label{fr-asym}
 \end{equation}
\begin{figure}
\centering\includegraphics[width=0.7\hsize]{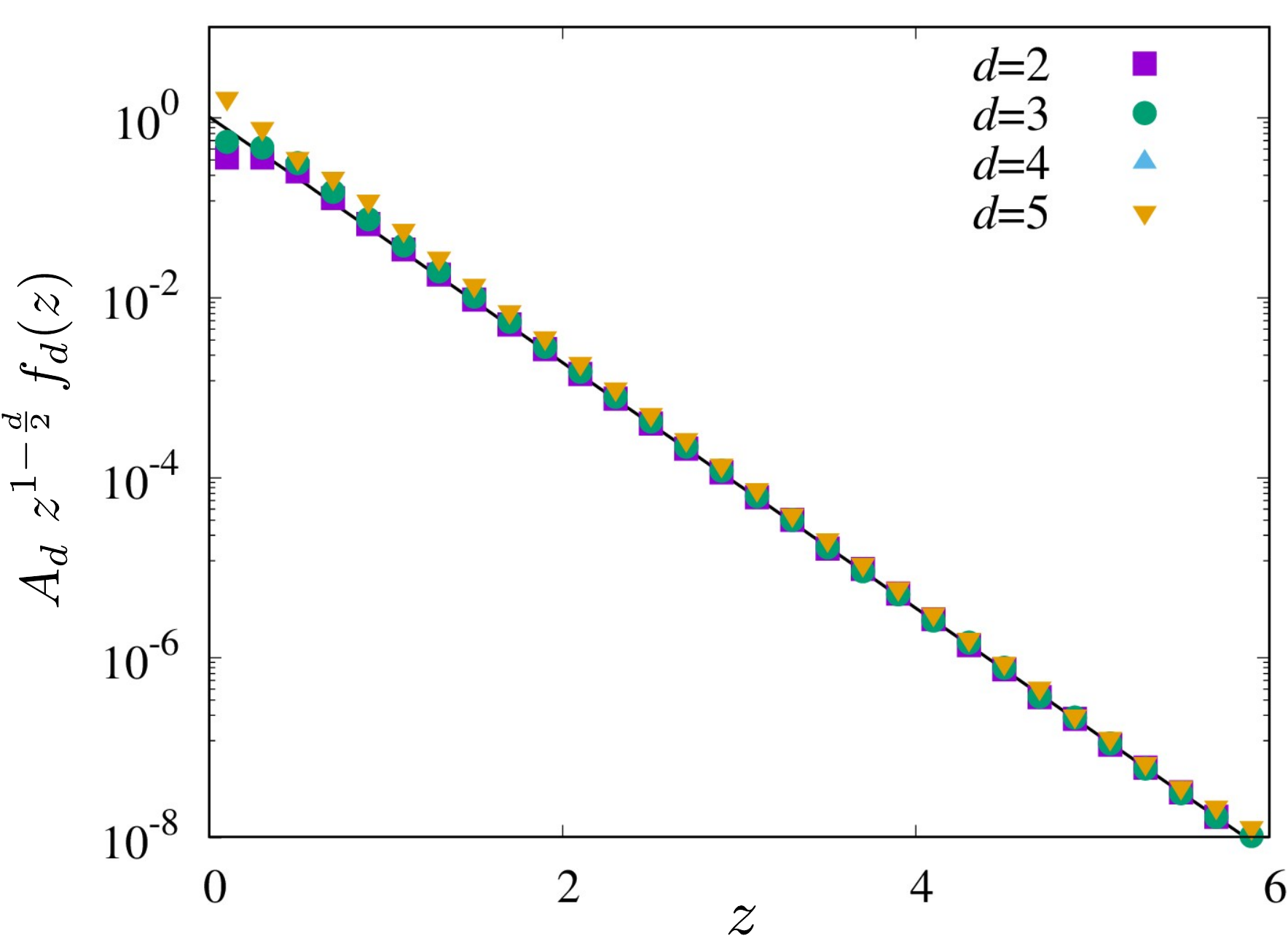}
\caption{Plot of the scaled marginal distribution for different dimensions predicted in \eref{fr-asym} (and $A_d=\Gamma(d/2)~2^{\frac{d-4}{2}}\pi^{\frac{1-d}{2}}$) with numerical simulations. The symbols denote the numerical simulations while the solid black line indicate the curve $e^{-\pi z}$.}
\label{f:dd_tails}
\end{figure} 
The universal tail behavior \eref{fr-asym} of the scaling function in arbitrary dimensions is compared with numerical simulations in 
\fref{f:dd_tails}, which shows a very good match, thus validating our prediction.

In the remaining part of the paper, the process studied above is subjected to stochastic resetting, where both the position and the diffusion coefficient are intermittently reset to their initial values at a constant rate. Under this resetting protocol, the position distribution reaches a stationary state in the $t\to\infty$. 
In the following sections, we study the stationary position distribution as well as the approach to the stationary state.

\section{Stationary position distribution in one dimension under stochastic resetting}\label{sec:R1d}
We start by defining the resetting protocol for the $d$-dimensional process. At a constant rate $\alpha$, the position components and the diffusion coefficient are reset to their initial values, which we take to be zero for the sake of simplicity. The introduction of resetting dynamics modifies the Langevin equations~\eref{ddim:Langevin} and \eref{model:intro} to, 
\begin{align}
x_i(t+dt)&=\begin{cases}x_i(t)+\sqrt{2D(t)\, dt}\,\tilde\eta_i&\quad\text{with probability }(1-\alpha\, dt)\\
0&\quad\text{with probability }\alpha\, dt,
\end{cases}\\
\intertext{and}
\omega(t+dt)&=\begin{cases}\omega(t)+\sqrt{2\Lambda^2\, dt}\,\tilde{\zeta}&\quad\text{with probability }(1-\alpha\, dt)\\
0&\quad\text{with probability }\alpha\, dt,
\end{cases}
\label{reset:dynamics}
\end{align}
where $\tilde\eta_i$ and $\tilde{\zeta}$ are independent Gaussian random numbers with zero mean and unit variance.

Let us first consider the one dimensional case. In the presence of resetting, it is straightforward to write the last renewal equation for the joint probability distribution $\mathsf{P}_{\alpha}(x,\omega,t)$,
\begin{align}
\mathsf{P}_{\alpha}(x,\omega,t)=e^{-\alpha t} \,\mathsf{P}(x,\omega,t)+\alpha\int_0^t ds\, e^{-\alpha s}\, \mathsf{P}(x,\omega,s),
\label{renew:joint}
\end{align}
where $\mathsf{P}(x,\omega,t)$ denotes the joint distribution in the absence of resetting [see \eref{fpeq}]. The first term on the right hand side corresponds to the contributions coming from the trajectories which have not undergone any resetting event until time $t$, while the second term combines the contributions from all the trajectories where the last resetting event occurred at time $t-s$. An equivalent renewal equation for the  position distribution $P_{\alpha}(x,t)$ is immediately obtained by integrating \eref{renew:joint} over $\omega$, 
\begin{align}
P_{\alpha}(x,t)=e^{-\alpha t}P(x,t)+\alpha\int_0^t ds\, e^{-\alpha s} P(x,s),
\label{renew:x} 
\end{align}
where $P(x,s)$ is the position distribution without resetting, given by \eref{pxt_scaled}.

 The repeated returns of the particle to the origin eventually lead to a stationary position distribution which is obtained by taking the $t\to\infty$ limit in~\eref{renew:x}. It is clear from \eref{renew:x} that in this limit, the first term on the right hand side vanishes, and we have the stationary position distribution,
\begin{align}
P^s_{\alpha}(x)=&\alpha\int_0^{\infty} ds\, e^{-\alpha s} P(x,s)=\frac{\alpha}{4\Lambda}\int_0^{\infty}\frac{ds}{s}\,e^{-\alpha s}\,f\left(\frac{x}{4\Lambda s}\right),
\label{renew:ssx} 
\end{align}
where we have used \eref{pxt_scaled} in the second step.
The above integral has the scaling form,
\begin{align}
P^s_{\alpha}(x)=\frac{\alpha}{4\Lambda}G\Big(\frac{\alpha x}{4\Lambda}\Big)
\label{res_sccform}
\end{align}
where the scaling function is,
\begin{align}
G(w)=\int_0^{\infty}\frac{dz}{z} e^{-|w|/z}f(z).
\label{res_sc:1d}
\end{align}
The above integral cannot be evaluated in closed form for the functional form of $f(z)$ given in \eref{scaled-propagator}. However, the asymptotic behaviors of  the scaled stationary position distribution $G(w)$ can be extracted using the limiting behaviors of $f(z)$ for $z\to\infty$ and near $z\to 0$.
\subsection{Asymptotic distribution at the tails $|w|\gg 1$}
Let us first investigate how the scaled stationary position distribution behaves near the tails, i.e, for $w\gg 1$. Clearly, for large $w$, the dominant contribution to the integral in \eref{res_sc:1d} comes from large values of $z\sim w$. Consequently, using the asymptotic behavior of $f(z)$ for $z\gg 1$ [see \eref{free:largey}], we get,
\begin{align}
G(w)\approx\sqrt{\frac{2}{\pi}}\int_0^{\infty}\frac{dz}{z^{3/2}} e^{-|w|/z}e^{-\pi z}=\frac{e^{-2\sqrt{\pi |w|}}}{\sqrt{|w|/2}}\quad\text{for}~w\gg1.
\label{1dtail_sc}
\end{align}
Thus, the tails of the stationary distribution decay as a stretched exponential. This theoretical prediction agrees extremely well with numerical simulations, which is shown in the left panel of \fref{f:1dstationary}.
It is interesting to note that the stretched exponential decay obtained in \eref{1dtail_sc} is slower than the pure exponential decays typically observed in diffusion-like systems under resetting~\cite{evans2011diffusion,gupta2019stochastic,
santra2020run}. 

\subsection{Distribution near the origin $w\to 0$}

\begin{figure}
\centering\includegraphics[width=0.6\hsize]{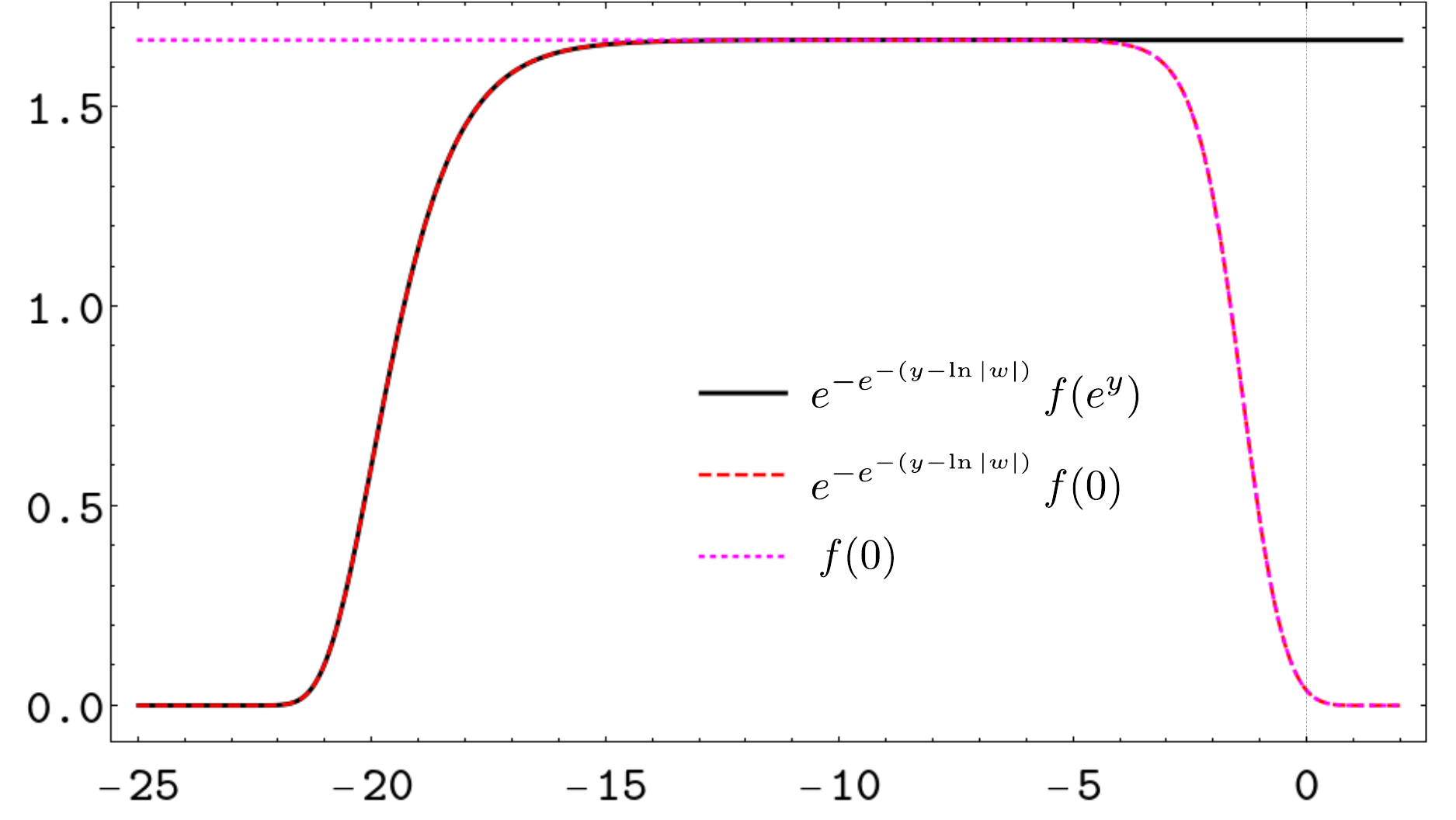}
\caption{Comparison of the  exact [eqn \eref{originexact}] and approximate integrand [eqn \eref{res:origin1d1}], in solid black line and red dashed line respectively, used to evaluate the stationary distribution near the origin in one dimension.}\label{f:origin_int}
\end{figure}

Next we focus on the behavior of the stationary position distribution near the resetting point, i.e., the origin $w=0$. This can be obtained systematically [see \ref{app:origin} for the details] using the integral form of $f(z)$ given in~\eref{characteristic-function} as, 
\begin{align}
G(w)=&-f(0)\ln |w|+\frac{b_1}{2}+\frac{c_1}{2}w+O(w^2) \quad\text{as}~w\to 0,
\label{1d:originf}
\end{align}
where the coefficients $b_1$ and $c_1$ are given in \ref{app:h1_abc}. 

The leading logarithmic divergence can be understood from the following heuristic argument. The contribution to the integral in \eref{res_sc:1d} for the region  $|w|\to 0$ comes mostly from the region $z\sim (O(w),1)$ because of the essential  singularity at $z=0$ and fast decay of $f(z)$ for $z\gg 1$. It is best visualized by making the change of variable $z=e^y$ in \eref{res_sc:1d}, which gives,
\begin{align}
G(w)=\int_{-\infty}^{\infty} dy \exp\Big[-e^{-(y-\ln|w|)}\Big]f(e^y).\label{originexact}
\end{align}
The function $f(e^y)$ decays fast to zero for $y>0$. On the other hand, $f(e^y)$  increases sharply and tends to the constant value $f(0)$ as $y$ decreases from zero to $-\infty$ [see \fref{f:origin_int}]. Therefore, the dominant part of \eref{originexact} can be obtained from,
\begin{align}
G(w)\approx&\int_{-\infty}^{0} dy\exp\Big[-e^{-(y-\ln|w|)}\Big]f(0)= - f(0)\ln |w|+O(1).
\label{res:origin1d1}
\end{align}

\begin{figure}
\includegraphics[width=\hsize]{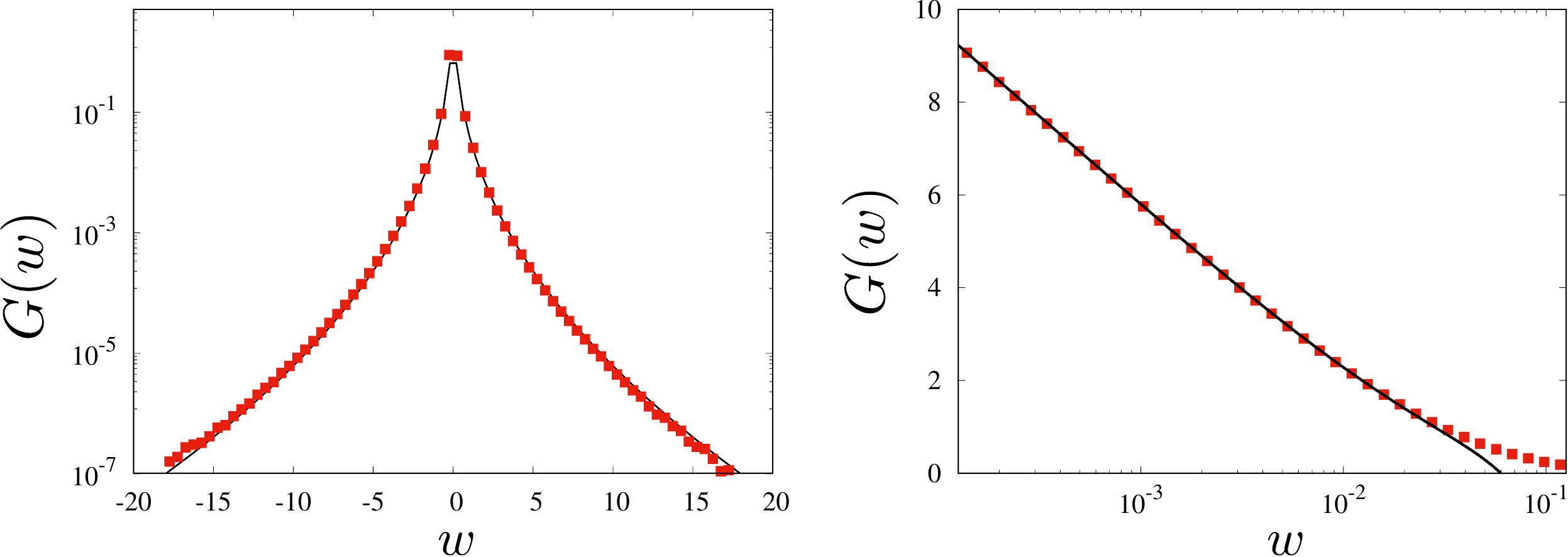}
\caption{Plot of the scaled stationary position distribution in one dimension. Left panel shows the behavior at the tails while the right panel shows the logarithmic divergence at the origin. The symbols denote numerical simulations while the solid black lines correspond to the predicted analytical forms \eref{1dtail_sc} and \eref{1d:originf} respectively.}
\label{f:1dstationary}
\end{figure}

 A comparison of the analytical prediction \eref{1d:originf} with the numerical simulations in \fref{f:1dstationary} (right panel) shows an excellent agreement.
It is worth mentioning that the logarithmic divergence of the position distribution at the origin is in stark contrast to the behavior of ordinary Brownian motion under stochastic resetting, where the probability density of finding the particle at the resetting position is finite. 

%

\section{Stationary position distribution in $d$-dimensions under stochastic resetting}\label{sec:Rdd}
Let us now consider the general $d$-dimensional case with the resetting dynamics~\eref{reset:dynamics}. We focus on the radial distribution, since the resetting to the initial position does not affect the isotropy of the distribution. The renewal equation \eref{renew:x} for the one-dimensional case generalizes to,
\begin{align}
P_\alpha(r,t)=e^{-\alpha t}P(r,t)+\alpha\int_0^{t}ds \,e^{-\alpha s}P(r,s),
\end{align}
where $P_\alpha(r,t)$ is the $d$-dimensional radial position distribution in the presence of resetting and $P(r,t)$ is the radial distribution in the absence of resetting given by \eref{radialfree}.

The stationary distribution $P^s_\alpha(r)$ is obtained by taking the $t\to\infty$ limit of the above equation,
\begin{align}
P^s_\alpha(r)=\alpha\int_0^{\infty}ds \,e^{-\alpha s}P(r,s)=\frac{\alpha}{4\Lambda}\int_0^\infty\frac{ds}{s}e^{-\alpha s}f_d\left(\frac{r}{4\Lambda s}\right),
\end{align}
where $f_d(z)$ is given by Eq.~\eqref{f_d-r2}. It can be immediately seen that the above integral has the scaling form,
\begin{align}
P^s_\alpha(r)=\frac{\alpha}{4\Lambda}H_d\Big(\frac{\alpha r}{4\Lambda} \Big),
\label{radialreset_us}
\end{align}
where the scaling function is given by 
\begin{align}
H_d(w)=\int_0^{\infty}\frac{dz}{z} e^{-w/z}f_d(z).
\label{res_sc:ddim}
\end{align}
It is important to note that $H_1(w)$ is related to the scaled one-dimensional position distribution $G(w)$ by the simple relation,
\begin{align}
G(w)=H_1(w)/2.\label{g(w)full}
\end{align}
This can be obtained by writing \eref{res_sc:ddim} using 
\eref{f_d-r2} for $d=1$, 
\begin{align}
H_1(w)=\sqrt{\frac{2}{\pi}}\int_0^{\infty}dz\, z^{-\frac 12} e^{-w/z} M_{-\frac 12}(z)=2G(w),
\end{align}
where we used the explicit form $M_{-\frac 12}(z)=\sqrt{\frac2\pi}\,z^{-1/2}\cos z$ to arrive at the second equality.

Using \eref{f_d-r2} in \eref{res_sc:ddim}, in the above equation, we get,
\begin{align}
H_d(w)=L_d\int_0^\infty dz~ z^{d-2}e^{-w/z}\Bigg(-\frac{1}{z}\frac{d}{dz}\Bigg)^nF(z)\quad 
\text{with }F(z)=z^{-\beta}M_\beta(z),
\label{hdw0}
\end{align}
where $L_d=2^{2-\frac d2}/\Gamma(d/2)$ and $n=(d-2\beta)/2-1$ (with $\beta=-1/2,\,0$ for odd and even dimensions respectively) is an integer.

The above equation allows us to express the scaled radial distribution for $d>2$ in terms of the $H_1(w)$ and $H_2(w)$ for odd and even dimensions, respectively, as we show below.
 It follows from \eref{f_d-r2} and \eref{fr-asym} that for $d=1$ and $2$, $f_d(z)=L_d z^{d-1} F(z)\sim e^{-\pi z}$ as $z\to\infty$.
Performing an integration by parts on the right hand side of \eref{hdw0} and noting that the boundary terms vanish, we get,
\begin{align}
H_d(w)=&L_d\int_0^\infty \Big[(d-3)z^{d-4}+wz^{d-5}\Big]\,e^{-w/z}\Bigg(-\frac{1}{z}\frac{d}{dz}\Bigg)^{n-1}F(z).
\label{hdw1}
\end{align}
Comparing the right hand sides of \eref{hdw1} and  \eref{hdw0}, and using $L_d/L_{d-2}=1/(d-2)$, we find the recursive equation,
\begin{align}
H_d(w)=&\frac{1}{d-2}\Big[(d-3)H_{d-2}(w)-wH_{d-2}'(w)\Big].
\label{H_drecur}
\end{align}
Clearly, it suffices to obtain $H_1(w)$ and $H_2(w)$, as the above recursion relation allows us to express $H_d(w)$ eventually in terms of $H_1(w)$ and $H_2(w)$ respectively for odd and even dimensions. Unfortunately, even for $d=1$ and $d=2$, the integrals in \eref{res_sc:ddim} cannot be evaluated exactly to yield a closed-form expression. However, the behaviors near the origin and at the tail can be extracted, as we see below.

\subsection{Distribution near the origin $w\to 0$}

Let us first look at the radial distribution near the origin. Using  \eref{f_d-r} in \eref{res_sc:ddim} and interchanging the order of the integrations, we get, 
\begin{align}
H_d(w)=\frac{2^{1-\frac d2}}{\Gamma(d/2)}\int_0^{\infty}
d\kappa \frac{1}{\sqrt{\cosh(\kappa/2)}}\int_0^{\infty} dy~y^{\frac{d}{2}-1}\,e^{-w \kappa/y}J_{\frac d2-1}(y).
\end{align}
The integral over $y$ can be computed exactly for $d=1$ and $2$ and yields,
\begin{align}
H_1(w)=&\frac{1}{\pi}\int_0^{\infty} d\kappa\frac{1}{\sqrt{\cosh(\kappa/2)}}\mathcal{G}^{3\,0}_{0\, 4}
\Bigg(\frac{w^2\kappa^2}{16}\Bigg|
\begin{array}{c}
 0,0,\frac{1}{2},\frac{1}{2} \label{h1_int}\\
\end{array}
\Bigg),
\intertext{and}
H_2(w)=&\frac{1}{4\sqrt{\pi}}\int_0^\infty d\kappa\frac{1}{\sqrt{\cosh(\kappa/2)}}\mathcal{G}^{3\,0}_{0\, 4}\Bigg(\frac{w^2\kappa^2}{16}\Bigg| -\frac 12,0,0,0 \Bigg),\label{h2_int}
\end{align}
where $\mathcal{G}$ denotes the Meijer-G functions~\cite{dlmf}.
The integrals over $\kappa$ do not yield any closed form expressions. However, to obtain the behavior near $w=0$, we can expand the integrand near $w=0$, and then integrate term by term. This leads to the small $w$ behavior for $H_1(w)$ and $H_2(w)$ as,
\begin{align}
H_1(w)&=-a_1\ln w+b_1+c_1w+d_1w^2+e_1w^2\ln w+h_1w^3+O(w^4)
\\ 
\intertext{and}
H_2(w)&=a_2+b_2 w+c_2 w \log (w)+d_2 w^2+e_2 w^3+h_2 w^3 \log (w)+O(w^4),
\end{align} 
where the leading coefficients $a_2$ and $a_1$ are given by,
\begin{align}
a_2=\int_{0}^{\infty}\frac{d\kappa}{\sqrt{\cosh(\kappa/2)}}=\frac{[\Gamma(1/4)]^2}{\sqrt{2\pi}}\quad\text{and}~~a_1=\frac{2 a_2}{\pi}.\label{a1a2}
\end{align}
The remaining coefficients  are given in \ref{app:origin} [eqns \eref{app:h1_abc} and \eref{app_h2abc}].

 Having obtained $H_1(w)$ and $H_2(w)$, we can now find the distribution for $d>2$ using the recursion relation \eref{H_drecur}. For example, the explicit leading behaviors of the scaled radial distributions near the origin for $d=3,~4,$ and $5$ are given by, 
\begin{subequations}
\begin{align}
H_3(w)&=a_1-c_1 w-2 d_1 w^2-e_1 w^2-2 e_1 w^2 \log (w)-3h_1w^3+O(w^4),\label{h3}\\
H_4(w)&=\frac 12(a_2-c_2 w-d_2 w^2-2 e_2 w^3-h_2 w^3-2 h_2 w^3 \ln w)+O(w^4)\label{h4},\\
H_5(w)&=\frac{1}{3}(2 a_1+c_1 w-2 e_1 w^2-3h_1 w^3)+O(w^4)\label{h5}.
\end{align}
\label{hdorigin}
\end{subequations}
The different coefficients appearing in the above equations are provided in \ref{app:origin} [see eqns \eref{app:h1_abc} and \eref{app_h2abc}].
 We compare the above results with numerical simulations in the right panel of~\fref{f:hd_reset} and find good agreement for small $w$. Thus for $d>2$. the stationary radial distribution has a finite value near the resetting point. Interestingly, for a standard diffusing particle with resetting, the distribution vanishes near the origin, in stark contrast to what we see above.

 Note that, the logarithmic divergence at the origin seen for $d=1$ disappears for $d>1$, where the distribution approaches a $d$-dependent constant as $w\to 0$. For example $H_3(0)=a_1$, $H_4(0)=a_2/2$ and so on. In fact, the constant $H_d(0)$ for $d\geq4$ can be obtained by taking $w\to 0$ in~\eref{H_drecur} as,
\begin{align}
H_d(0)=\Bigg(\frac{d-3}{d-2}\Bigg)H_{d-2}(0)=\begin{cases}\displaystyle
\frac{(d-3)!!}{2^{\frac{d-2}{2}}(\frac{d-2}{2})!}\,a_2\quad&\text{for even }d\geq4,\\[2em]
\displaystyle
\frac{2^{\frac{d-3}{2}}(\frac{d-3}{2})!}{(d-2)!!}\,a_1\quad&\text{for odd }d\geq 5,
\end{cases}
\end{align}
where $a_1$ and $a_2$ are given by \eref{a1a2}.

\subsection{Asymptotic distribution at the tail $w\gg 1$}

\begin{figure}
\includegraphics[width=\hsize]{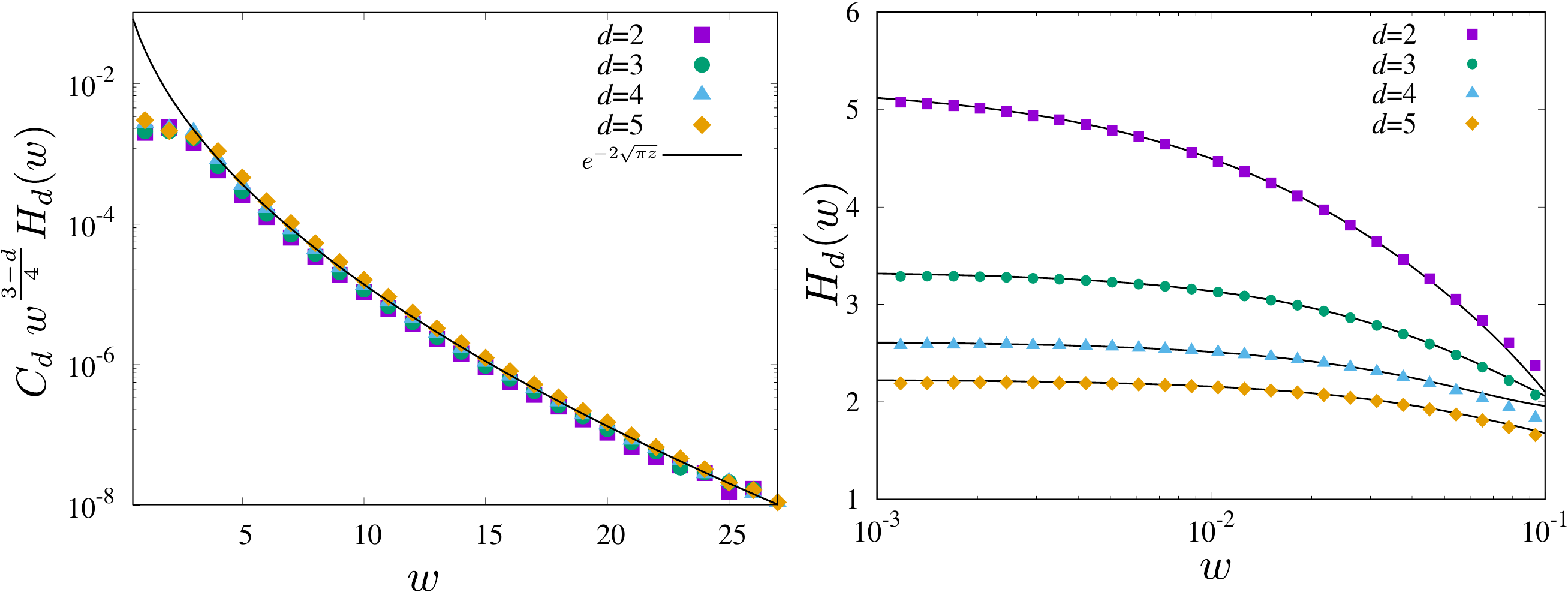}
\caption{Plot of the scaled stationary radial distribution, $H_d(w)$ defined in \eref{radialreset_us}, for different dimensions. Left panel compares the distribution obtained from numerical simulation (symbols) with the analytical prediction \eref{tails_ddim} (solid black line). Here, $C_d=2^{\frac  {d -4}{2}}\pi^{-\frac{d+1}{4}}\Gamma(\frac d2)$ denotes a $w$-independent scaling factor. The right panel magnifies the region near the origin $w=0$ to compare the analytical prediction \eref{hdorigin} for small $w$ (solid black lines) with numerical simulations (symbols).}
\label{f:hd_reset}
\end{figure}

The asymptotic distribution at the tail for $d>1$ shows the same stretched exponential decay as in $d=1$, as we will see below.
 From \eref{res_sc:ddim}, it can be seen that the behavior of the scaling function $H_d(w)$ for large $w$, is dominated by the large $z$ behavior of $f_d(z).$ Using the universal tail behavior of $f_d(z)$ given by \eqref{fr-asym}, we get, for $w\gg 1$,
\begin{align}
H_d(w)\approx{4\pi^{d/2} \over 2^{d/2}\sqrt{\pi} \Gamma(d/2)}\int_0^\infty dz\,z^{(d-4)/2}e^{-w/z}e^{-\pi z}.		
\end{align}
Performing the integral, we get the scaling function at the tail as,
\begin{align}
H_d(w)\approx&\frac{2^{3-\frac{d}{2}} \pi ^{d/4} }{\Gamma \left(\frac{d}{2}\right)}w^{\frac{d-2}{4}}\,K_{1-d/2}\left(2\sqrt{\pi w}\right)\label{tails_bessel}\\
=&\frac{2^{2-\frac{d}{2}} \pi ^{\frac{d+1}{4}} }{\Gamma \left(\frac{d}{2}\right)}w^{\frac{d-3}{4}}\,e^{-2\sqrt{\pi w}}\Big[1+O(w^{-\frac{1}{2}})\Big],
\label{tails_ddim}
\end{align}
where the second line is obtained from the asymptotic behavior of the modified Bessel function of the second kind $K_\nu(z)$ for large $z$. Note that the predicted universal stretched exponential decay is same as in $d=1$ [see \eref{1dtail_sc}]; this is in contrast to the behavior near the origin, which is drastically different for $d=1$ and $d>1$. The analytical prediction \eref{tails_ddim} is validated in \fref{f:hd_reset} (left panel) using numerical simulations for $d=2,3,4$ and $5$.

\section{Approach to the stationary states}\label{s:approach}


In this section, we focus on the dynamical behavior of the position distribution of the process given by~\eref{intro1} in the presence of resetting. In particular, we investigate the temporal relaxation to the stationary state~\eref{res_sccform} in $d=1$ and show that it exhibits a dynamical transition similar to the one studied in~\cite{majumdar2015dynamical}---as time progresses, the particle position attains a stationary state in an inner core region around the resetting position, while the region outside the inner core remains transient. Interestingly, the stationary inner core region grows with a constant acceleration with time as opposed to a constant velocity growth for standard diffusion, as we will show below.

We start by rewriting the renewal equation \eref{renew:x} in terms of the scaling form \eref{pxt_scaled}, 
\begin{align}
P_{\alpha}(x,t)=\frac{\alpha}{4\Lambda t}e^{-\alpha t}f\Big(\frac{x}{4\Lambda t}\Big)+\frac{\alpha}{4\Lambda}\int_{\frac{|x|}{4\Lambda t}}^{\infty}\frac{dz}{z}e^{-\frac{\alpha|x|}{4\Lambda z}}f(z),
\label{fullrenewaleq}
\end{align}
 where $f(z)$ is given by \eref{scaled-propagator}. 
 For sufficiently large $t\gg\alpha^{-1}$ and $|x|\ll 4\Lambda t$,  the first term vanishes exponentially and the lower limit of the integration in the second term approaches zero. Thus, in this inner core $P_{\alpha}(x,t)$ converges to   $P_{\alpha}^s(x)$ given in \eref{renew:ssx}.
  
  To understand the behavior for $|x|\gg 4\Lambda t$, we note from \eref{fullrenewaleq} that only the large $z$ form $f(z)\sim \sqrt{\frac 2\pi}\,  |z|^{-1/2}\,e^{-\pi |z|}$  [see \eref{free:largey}]
   is relevant for the integral. It is convenient to rewrite  \eref{fullrenewaleq}, by making a change of variable $z=|x|/(4\Lambda\tau t)$ as,
  \begin{align}
P_{\alpha}(x,t)\approx \frac{\alpha }{\sqrt{2\Lambda \pi t}}\frac{e^{-t\phi(1,y)}}{\sqrt{|x|}}
+\frac{\alpha\sqrt{t}}{\sqrt{2\pi\Lambda}}\int_0^1\frac{d\tau}{\sqrt\tau}\frac{e^{-t\phi(\tau,y)}}{\sqrt{|x|}}\quad\text{with}~y=\frac{\pi |x|}{4\Lambda t^2},
 \label{resetfull2}
\end{align}
where $\phi(\tau,y)=\alpha\tau+y/\tau$. At very large values
of $t$ and with a ﬁxed $y$, we can estimate the integral using the lowest value  of $\phi(\tau,y)$ in $\tau\in[0,1]$. The function 
$\phi(\tau,y)$ is minimum at $\tau^*=\sqrt{y/\alpha}$, which is obtained from
 $\frac{\partial \phi(\tau,y)}{\partial\tau}|_{\tau=\tau^*}=0$.
Two different scenarios emerge depending on whether $\tau^*<1$ or $\tau^*>1$.

For $\tau^*<1$, the first term in  \eref{resetfull2} is negligible compared to the second term, where the dominant contribution comes from the saddle point, resulting in $P_{\alpha}(x=4\Lambda y t^2/\pi,t)\sim e^{-t\phi(\tau^*,y)}$ for $y<\alpha$.
On the other hand, for $\tau^*>1$, the minimum of $\phi(\tau^*,y)$ lies outside the range of integration $(0,1)$. Thus, the lowest value of $\phi(\tau,y)$ within the integration limit is at the boundary $\tau=1$. As a result, $P_{\alpha}(x=4\Lambda y t^2/\pi,t)\sim e^{-t\phi(1,y)}$ for $y>\alpha$.
Note that, in this case, both the terms in \eref{resetfull2} are of the same order, implying that the contribution to the probability distribution comes from trajectories that have undergone none or very few resettings. 

\begin{figure}
\includegraphics[width=\hsize]{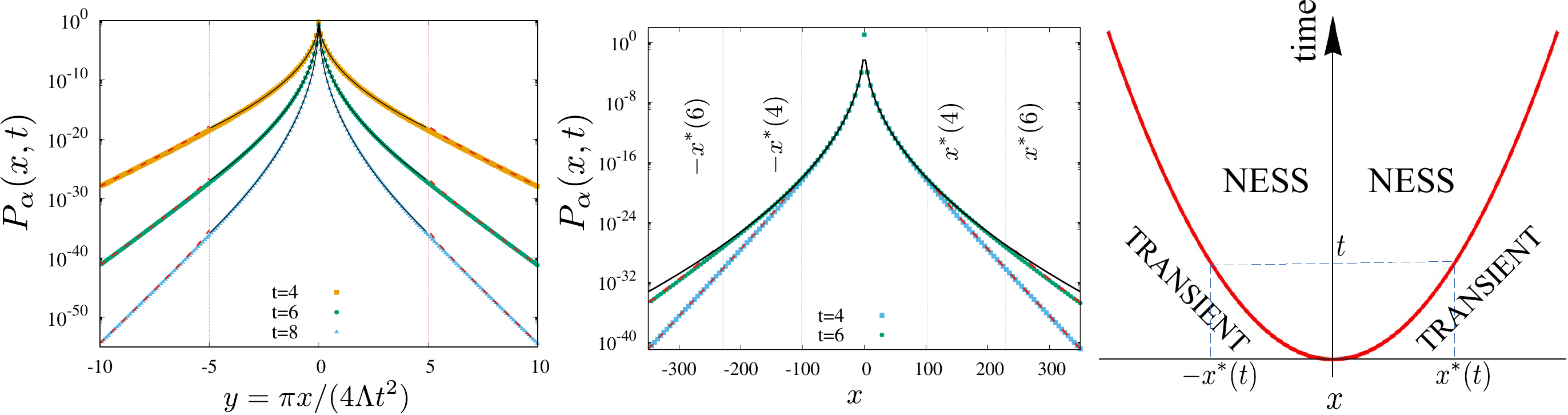}
\caption{Approach to the stationary state in one dimension. The left panel shows the position distribution against the scaled variable $y=\frac{\pi x}{4\Lambda t^2}$ at three different times---the red dotted line, denoting $y=\alpha$, separates the inner and the outer regions predicted in \eref{ytrans}; the symbols denote the exact distribution obtained by numerical integration of \eref{fullrenewaleq}, the solid black line and red dashed lines denote the analytical predictions for $y<\alpha$ and $y>\alpha$ in \eref{ytrans} respectively. The middle panel shows the distribution as a function of $x$---the symbols denote the exact distribution obtained by numerical integration of \eref{fullrenewaleq}, while the solid black line and the red dashed lines correspond to the stationary distribution and transient distributions predicted in \eref{finite-t}; the dotted vertical lines denote the boundary of the inner stationary, given by $x^*(t)$. A schematic representation of the approach to the stationary state is illustrated in the right panel.}
\label{f:transient}
\end{figure}

Therefore, the position distribution has the following large deviation form,
\begin{align}
P_\alpha(x,t)\sim \exp\left[-t\, {\cal I}\left(\frac{\pi |x|}{4\Lambda t^2}\right)\right], \label{eq:Pxt_ldf}
\end{align}
 where the large deviation function,
\begin{align}
{\cal I}(y)=\begin{cases}2\sqrt{\alpha y}\qquad &\text{for }y<\alpha,\\
\alpha+y\qquad &\text{for }y>\alpha.
\end{cases}
\label{ytrans}
\end{align}
Note that, the large deviation function and its first derivative are continuous at $y=\alpha$, while the second derivative exhibits a discontinuity. This transition is illustrated in \fref{f:transient} (left panel) for three different times.
 Rewriting \eref{ytrans} in terms of the original variables $x,$ $t$, we get, from \eref{eq:Pxt_ldf},
\begin{align}
P_\alpha(x,t)\sim
\begin{cases}
\displaystyle e^{-\sqrt{\frac{\alpha\pi |x|}{\Lambda}}}\qquad &\text{for }|x|< x^*(t),\\
\displaystyle e^{-\alpha t-\frac{ \pi |x|}{4\Lambda t}}\qquad &\text{for }|x|> x^*(t),
\end{cases}\quad
\text{where}~x^*(t)=\frac{4\alpha\Lambda t^2}{\pi}.
\label{finite-t}
\end{align}
This is illustrated in \fref{f:transient} (middle panel) by comparing with $P_\alpha(x,t)$ obtained from exact numerical evaluation of \eref{resetfull2}. Note that the subleading contributions to \eref{finite-t} are calculated in the~\ref{app:ldf}.
 Thus we see that the boundary of the inner stationary region grows with a constant acceleration $8\alpha\Lambda/\pi$, in contrast to standard diffusion where it grows at a constant velocity. Figure~\ref{f:transient} (right panel) shows a schematic representation of this dynamical transition in the approach to the stationary state. 

\section{Summary and Conclusions}\label{concl}
We study a Brownian motion in arbitrary dimensions with a stochastically evolving diffusion constant. In one dimension, the position distribution has a ballistic scaling form with exponentially decaying tails. We further show that in arbitrary dimensions $d>1$,  the marginal radial distribution has the same ballistic scaling form accompanied by the same exponential tails. In particular, we find the exact scaling function for odd dimensions. In the presence of a resetting dynamics, which restarts the process from its initial condition at a constant rate, the position distribution in all dimensions reaches a stationary state with a universal stretched exponential tail. The stationary distribution at the origin shows a logarithmic divergence at the origin in $d=1$, while it approaches a $d$-dependent finite value for $d>1$. We also study the approach to this stationary state and find that at finite times, the inner region around the resetting position reaches the stationary state; this region grows with a constant acceleration proportional to the resetting rate. 

Interesting open questions include studying the effects of other resetting protocols, like resetting of only position or only diffusion coefficient reset, on this dynamics. Recently different non-instantaneous resetting protocols have been studied~\cite{santra2021brownian,gupta2020stochastic,radice2021one}, which opens up the possibility of experimental realizations. It would be interesting to see the effects of such non-instantaneous resetting protocols on the dynamics of a Brownian particle with a stochastically evolving diffusion coefficient.

\appendix

\section{Asymptotic tail behavior of the radial position distribution for odd and even dimensions}\label{App:A}
In this section we derive the asymptotic tail behavior of the radial position distribution announced in \eref{fr-asym} for  odd and even dimensions.

\subsection{Odd dimesnions}

We set $d=2n+1$ where $n=1,2,3,\dotsc$. Therefore, $d/2-1=n+\beta$ with $\beta=1/2$. Now using $J_{-1/2}(x)=\sqrt{2/\pi} \,\cos(x)/\sqrt{x}$, from \eref{G-alpha}, we get,
\begin{equation}
\sqrt{z}\,M_{-{1\over 2}} (z) =\sqrt{{2\over\pi }}
\int_0^\infty\frac{ \cos(\kappa z)}{\sqrt{\cosh(\kappa/2)}}\, d\kappa.
\end{equation}
The above integral can be performed explicitly,  which gives
\begin{equation}
\sqrt{z}\,M_{-{1\over 2}}(z)={1\over \pi}\, \Gamma\left({1\over 4} + i z\right)\,
\Gamma\left({1\over 4} - i z\right).
\end{equation} 
Using the asymptotic behavior $\sqrt{z}\, M_{-{1\over 2}}(z)\sim 2 e^{-\pi z}/\sqrt{z}$ for large $z$,  and writing $n=(d-1)/2$, we get the scaled radial distribution using \eref{ddim:oddexact},
\begin{equation}
f_d(z) \sim \frac{2\sqrt{\pi}}{\Gamma(d/2)}\left(\frac{\pi z}{2}\right)^{\frac d2-1}  \,e^{-\pi z}
\quad\text{for}~z\gg 1.
\label{fr-asym-odd}
\end{equation}

\subsection{Even dimensions}

Now we give the derivation of ~\eref{fr-asym} for even dimensions.
Let us start from \eref{G-alpha} in the main text. Setting $\beta=0$ for even dimensions, we have,
\begin{equation}
M_0(z)=\int_0^\infty\, \frac{J_0(\kappa z)}{\sqrt{\cosh (\kappa/2)}}\, \kappa\, d\kappa.
\label{Gr2}
\end{equation}
It is difficult to perform the integral exactly. However, the asymptotic behavior of $M_0(z)$ for large $z$, can be found by analyzing the singularities of the function $1/\sqrt{\cosh(\kappa/2)}$, which has branch point singularities at $\kappa_m=\pm i (2m+1)\pi$ with $m=0,1,2,\dotsc$. It is useful to use the identity 
\begin{equation}
{1\over \sqrt{\cosh(\kappa/2)}} = 2\left[\sum_{m=0}^\infty {(-1)^m (2m+1) \pi\over (2m+1)^2\pi^2+\kappa^2}\right]^{1/2}.
\end{equation}
Anticipating that the large $z$ behavior of $M_0(z)$ is dominated by the contributions from the singularities closest to the origin, we replace $1/\sqrt{\cosh(\kappa/2)}$  by the term $2\sqrt\pi/\sqrt{\kappa^2+\pi^2}$ corresponding to $m=0$ and perform the integral \eref{Gr2}.  This gives the leading asymptotic behavior as 
\begin{equation}
M_0(z) \sim {2\sqrt{\pi} \over z} \,e^{-\pi z}\quad\text{for}~z\gg 1.
\end{equation}
Therefore, from \eref{f_d-r2} the leading asymptotic behavior of $f_d(z)$  is given by,
 \begin{equation}
 f_d(z) \sim \frac{2\sqrt{\pi}}{\Gamma(d/2)}\left(\frac{\pi}{2}\right)^{\frac d2-1} \,  \,e^{-\pi z}
\quad\text{for}~z\gg 1,
\label{fr-asym-even}
 \end{equation}
which is the result \eref{fr-asym} quoted in the main text.

\section{Coefficients for the expansion of $H_1(w)$ and $H_2(w)$ in the region $w\ll 1$}\label{app:origin}
In this section we give the exact forms of the coefficients $\{a_i,b_i,c_i,d_i,e_i,g_i\}$ which occur in the series expansion of $H_d(w)$.
Expanding the Meijer-G function in \eref{h1_int} in a Taylor series about $w=0$ and integrating term by term, we get,
\begin{align}
H_1(w)=-a_1\ln w+b_1+c_1w+d_1w^2+e_1w^2\ln w+h_1w^3+O(w^4),
\label{app:H_1}
\end{align} 
which is quoted in the main text. Here,
\begin{subequations}
\begin{align}
a_1&=\frac{2}{\pi}\int_0^\infty d\kappa ~g(\kappa)
=2f(0)=\sqrt{\frac{2}{\pi^3}}\,[\Gamma(1/4)]^2=3.34\dotsc\\
b_1&=-\frac{2}{\pi}\int_0^\infty d\kappa~ (2E+\ln \kappa)~g(\kappa)=-2(Ef(0)+\int_0^\infty \frac{d\kappa}{\pi}~\ln \kappa~g(\kappa))=-6.93\dotsc\\
c_1&=\int_0^\infty d\kappa~ \kappa~ g(\kappa)=22.25\dotsc\dotsc\\
d_1&=\frac{1}{2\pi}\int_0^\infty d\kappa~\kappa^2(2E-3+\ln \kappa)~g(\kappa)=13.47\dotsc\\
e_1&=\frac{1}{2\pi}\int_0^\infty d\kappa~ \kappa^2~ g(\kappa)=28.71\dotsc\\
h_1&=-\frac{1}{36}\int_0^\infty d\kappa~ \kappa^3~ g(\kappa)=-60.29\dotsc
\end{align}
\label{app:h1_abc}
\end{subequations}
with $g(\kappa)=1/\sqrt{\cosh(\kappa/2)}$ and $E$ denotes the Euler-gamma constant.

Next for $H_2(w)$, expanding the Meijer-G function in \eref{h1_int} as a Taylor series about $w=0$ and integrating term by term, we get,
\begin{align}
H_2(w)=a_2+b_2 w+c_2 w \log (w)+d_2 w^2+e_2 w^3+h_2 w^3 \log (w)+O(w^4),
%
\end{align}
where the coefficients are given by,
\begin{subequations}
\begin{align}
a_2&=\int_0^\infty d\kappa~ g(\kappa)=\frac{\Gamma \left(\frac{1}{4}\right)^2}{\sqrt{2 \pi }}=5.24\dotsc\\
b_2&=\int_0^\infty d\kappa~ \kappa~g(\kappa)\Big(2E-1+\ln (\kappa/2)\Big)=28.83\dotsc\\
c_2&=\int_0^\infty d\kappa~ \kappa~g(\kappa)=22.25\dotsc\\
d_2&=-\frac{1}{2}\int_0^\infty d\kappa~ \kappa^2~g(\kappa)=-90.18\dotsc\\
e_2&=\frac{1}{48}\int_0^\infty d\kappa~ \kappa^3~g(\kappa)\Big(\frac{17}{3}-4E+2\ln2-2\ln \kappa\Big)=-24.56\dotsc\\
h_2&=\frac{-1}{24}\int_0^\infty d\kappa~ \kappa^3~g(\kappa)=-90.44\dotsc .
\end{align}
\label{app_h2abc}
\end{subequations}
The coefficients obtained above in \eref{app:h1_abc} and \eref{app_h2abc} also appear in the determination of $H_d(w)$ for $d>3$.

\section{Subleading terms of the finite time position distributions}\label{app:ldf}

In this section, we calculate the subleading terms of the probability disributions \eref{finite-t} for the stationary region ($\tau^*<1$) and the transient region ($\tau^*>1$).

\begin{itemize}
\item $\tau^*<1$: In this case the minimum of $\phi(\tau,y)$ lies within the limits of integration $(0,1)$. Thus, we can approximate $\phi(\tau,y)$ inside the integral by $\phi(\tau^*,y)+\frac{\tau^2}{2}\phi''(\tau^*,y)$ and thereafter doing the $\tau$ integral, we have,
\begin{align}
P_{\alpha}(x,t)\approx & \, \alpha\sqrt{\frac{2t}{4\Lambda\tau^* \pi |x|}}\frac{\sqrt{2\pi}\, \exp{\left[-t \,\phi(\tau^*,\frac{\pi |x|}{4\Lambda t^2})\right]}}{\sqrt{t\phi''(\tau^*,\frac{\pi |x|}{4\Lambda t^2})}} \n\\[0.5em] 
=&\, \sqrt{\frac{\alpha}{2\Lambda |x|}}\,\exp{\left[-\sqrt{\frac{\alpha\pi |x|}{\Lambda}}\right]}\equiv P_{\alpha}^s(x).
\end{align}
where $P_{\alpha}^s(x)$ is the stationary state distribution \eref{1dtail_sc}.
\item $\tau^*>1$: In this case the minimum of $\phi(\tau,y)$ lies outside the limits of integration $(0,1)$. Thus, the minimum value of $\phi(\tau,y)$ within the integration limit is attained at the boundary $\tau=1$. So we approximate $\phi(\tau,y)$ inside the integral by $\phi(1,y)+(1-\tau)|\phi'(1,y)|$ and thereafter doing the $\tau$ integral, we have,
\begin{align}
P_{\alpha}(x,t)\approx\alpha  {\cal A}\sqrt{\frac{2t}{4\Lambda \pi |x|}} \exp{\left[-t\phi\left(1,\frac{\pi |x|}{4\Lambda t^2}\right)\right]}
\end{align}
where $ {\cal A}=\int_0^1d\tau \exp{\left[-t(1-\tau)|\phi'\left(1,\frac{\pi |x|}{4\Lambda t^2} \right)| \right]}$. 
\end{itemize}

\section*{References}
\bibliography{ref}

\end{document}